\documentclass[11pt]{article}

\usepackage[iso-8859-7]{inputenc}
\usepackage{epsfig}
\usepackage{graphicx}

\usepackage{amsfonts}
\usepackage{amssymb}
\usepackage{amsthm}
\usepackage{amsmath}
\usepackage{multirow}

\newcommand{\newc}{\newcommand}
\newc{\ra}{\rightarrow}
\newc{\lra}{\leftrightarrow}
\newc{\ov}{\overline}
\newc{\pa}{\partial}

\newc{\be}{\begin{equation}}
\newc{\ee}{\end{equation}}

\newc{\ba}{\begin{eqnarray}}
\newc{\ea}{\end{eqnarray}}

\newc{\D}{\Delta}

\newc{\la}{\lambda}
\newc{\vt}\vartheta
\newc{\nn}{\nonumber}

\begin{document}
\thispagestyle{empty}

\vskip 2truecm
\vspace*{3cm}
\begin{center}
{\large{\bf
On the GUT scale of {\cal F}-Theory $SU(5)$}}

\vspace*{1cm}

{\bf G.K. Leontaris$^{\,1}$, N.D. Vlachos$^{\,2}$}\\[3.5mm]
$^1\,$Theoretical Physics Division, Ioannina University, GR-45110 Ioannina, Greece \\[2.5mm]
$^2\,$Theoretical Physics
Division, Aristotle University, GR-54124 Thessaloniki, Greece

\end{center}

\vspace*{1cm}
\begin{center}
{\bf Abstract}
\end{center}

\noindent

In F-theory GUTs, threshold corrections from Kaluza-Klein (KK) massive modes arising from
gauge and matter multiplets  play an important role in the determination of the weak mixing angle
and the strong gauge coupling of the effective low energy  model. In this letter we further explore
 the induced modifications  on  the gauge couplings running and the GUT scale.
 In particular, we focus on the KK-contributions from  matter curves and analyse the conditions
 on the chiral and Higgs matter spectrum which imply a GUT scale consistent with the minimal
 unification scenario. As an application, we present an   explicit computation  of
 these thresholds for matter fields residing on  specific non-trivial Riemann  surfaces.

\vfill
\newpage

\section{Introduction}

It is well known that the spectrum of the minimal supersymmetric extension
of the Standard Model (SM) is consistent with a
 gauge couplings unification at a scale $M_{GUT}\sim 2\times 10^{16}$ GeV. This fact corroborates
the point of view that the SM gauge group factors emanate from a higher unified gauge symmetry.
In the simplest case, the SM gauge symmetry is embedded in the $SU(5)$  Grand unified Theory (GUT)  while
the SM  matter content is nicely   assembled into $SU(5)$ multiplets.
In addition, although string theory appears to be the appropriate candidate for incorporating gravity
into the unification scenario, one must  still confront  the
mismatch  between $M_{GUT}$ and the natural gravitational scale
 $M_{Pl}\sim 1.2 \times 10^{19}$ GeV.  Thus, a plausible implementation
 of unification,   requires a string theory formulation
 in which the gauge theory decouples from gravity at the desired scale.

Recently, there have been considerable efforts  to develop a viable effective
field theory model from F-theory~\cite{Vafa:1996xn}\footnote{For
comprehensive reviews see~\cite{Denef:2008wq,Weigand:2010wm,Heckman:2010bq}}.
 This picture consists of a 7-brane  wrapping a compact K\"ahler surface $S$  of two complex dimensions
 while the gauge theory of a particular model is associated with the geometric singularity of
the internal space~\cite{Bershadsky:1996nh,Donagi:2008ca,Beasley:2008dc,Beasley:2008kw}.
In this set up it is possible to decouple  gauge dynamics from gravity by restricting
to compact surfaces $S$ that are of del Pezzo type. The exact determination of the
GUT scale however, may depend on the spectrum and other details of the chosen
gauge symmetry and on the particular model.   In the present work, we will assume the minimal unified $SU(5)$ GUT.

  A reliable computation of the GUT scale should also take into consideration the various  threshold corrections.
  These may also depend on the choice of the  specific  compactification as well as the particular model. In F-theory
  $SU(5)$ we are examining here, there are several sources of threshold effects  that have to be taken into account~\cite{Donagi:2008kj,Blumenhagen:2008aw,Palti:2009,Leontaris:2009wi,Leontaris:2011pu,Heckman:2011hu}.
Thus, we encounter thresholds related  to the flux mechanism (used to
break the GUT gauge symmetry) which  induce splitting of the gauge
couplings at the GUT scale~\cite{Donagi:2008kj,Blumenhagen:2008aw}.
A second source concerns threshold corrections generated from heavy
KK massive modes~\cite{Donagi:2008kj,Leontaris:2011pu}.  Furthermore,  corrections
to gauge coupling running  arise due to the
appearance of  probe D3-branes  generically present in F-theory compactifations
and filling  the $3+1$ non-compact dimensions while sitting  at certain points
of the internal manifold~\cite{Heckman:2011hu}. Finally, threshold effects are
generated at scales $\mu <M_{GUT}$ when additional light degrees of
freedom and in particular superpartners are integrated out. The effects of
the latter have been extensively studied in the context of supersymmetric and
String Grand Unified Theories~\footnote{For an incomplete list see~\cite{Langacker:1992rq}.}.
In reasonable circumstances, (for example when no-extra degrees of freedom remain below $M_{GUT}$)
the last two categories can be made consistent with two loop corrections and
a unification scale of the order of $M_{GUT}\sim 2\times 10^{16}$GeV.

Thresholds induced by the flux mechanism have been extensively analyzed
in recent literature~\cite{Donagi:2008kj,Blumenhagen:2008aw,Leontaris:2009wi}.
 There, it   was shown that the $U(1)_Y$-flux induced splitting is compatible with the
 GUT embedding of the minimal supersymmetric standard model, provided that no
 extra matter other than  color triplets is present in the spectrum.
Thresholds originating
from KK-massive modes have been discussed in~\cite{Donagi:2008kj} and
were found to be related to a topologically invariant quantity, the Ray-Singer
analytic torsion~\cite{Ray:1973sb}.  This observation was originally
made for the case of manifolds with $G_2$ holonomy where thresholds
 were  computed  and estimates for the GUT scale were given~\cite{Friedmann:2002ty}.
In   the case of F-theory however, the situation is a little more complicated.
Indeed, in M-theory one assumes that massless $SU(5)$ multiplets are generated
at singularities of the internal space which are believed to be conical~\cite{Friedmann:2002ty}.
Since conical singularities induce no new length, it  is  expected that no new massive
particles are introduced.  On the contrary, in F-theory,
 KK-massive modes exist for both the gauge and the matter fields.
To be more precise, in the present context of the $SU(5)$ theory, these come
along  with  massless gauge fields propagating in the bulk, while the chiral matter
as well as the Higgs representations reside on two-dimensional Riemann surfaces  (matter curves).
In general, both kinds of  KK-modes contribute to the gauge coupling running and can
in principle modify the unification scale. It is straightforward to
estimate the modification induced by  the  vector supermultiplet, nevertheless
the contributions of the matter fields might be model dependent. In this letter
we aim to revisit this second source of threshold corrections.  We will discuss
this issue in the context of  models where chiral matter and
Higgs fields  occupy complete  $SU(5)$ multiplets. We will show that
under  reasonable assumptions for the matter curve bundle structure, no further
modifications are induced from the  corresponding matter  KK-massive modes.

To make the presentation self-contained,  we will first briefly review
the eight-dimensional twisted theory and obtain the degrees of freedom together with their
corresponding topological properties. In section 3 we will compute the threshold corrections. After
recapitulating  the basic steps for the gauge contributions, we will continue
with a detailed determination of the  thresholds from the  matter curves.  Next,
we will proceed with an explicit calculation of the KK-massive modes thresholds
originating from chiral and  Higgs matter curves and show that their only net effect
amounts to a shift of the common gauge coupling at the GUT scale.
In section 4 we will present our conclusions.

\section{Twisted gauge theory and  degrees of freedom}

Before  proceeding to the computation of  the threshold corrections and  following~\cite{Beasley:2008dc,Beasley:2008kw}, we will first review the salient  features of the theory and summarize the properties
of the massless and  massive degrees of freedom respectively.
 F-theory is defined locally by the worldvolume of 7-branes of ADE-type singularity
 which for definiteness we assume to be $SU(5)$.   We will further assume that a
  $U(1)_Y$ flux is turned on on the 7-brane in order to break $SU(5)$ down to the Standard
 Model (SM). We consider a maximally Supersymmetric Yang-Mills (YM) theory in 10
 dimensions on $R^{9,1}$  with field content consisting of a ten dimensional vector
 ${\cal A}_I, (I=0,1,\dots 9)$ and an adjoint valued fermion transforming under
 $SO(9,1)$ as a positive chirality  spinor representation of ${\bf 16}_+$.
 The supercharges are  also found to be in a ${\bf 16}_+$ representation.
 Under the reduction of the $R^{9,1}$ theory to $R^{7,1}$, the global symmetry
 of the resulting 8-dimensional theory reduces to
 \ba
 SO(9,1)&\ra& SO(7,1)\times U(1)_R\label{10r}\cdot
 \ea
 The 10-d gauge field ${\cal A}$ decomposes into an 8-d gauge field $A$ and two scalars
 $A_{8,9}$, combined into two complex fields
 \[\varphi=A_8+i A_9\in (1,+1),\;\bar\varphi=A_8-i A_9\in (1,-1)\]
which transform trivially under $SO(7,1)$ and have $\pm 1$ charges under $U(1)_R$
in (\ref{10r}). Also, from the spinor decomposition we get two chiral fermions
$\Psi_{\pm}$ transforming as
 \[{\bf 16}_+\ra (S_+,{\small\frac 12})+(S_-,-\frac 12) \cdot \]
Thereupon, the 8-d theory is compactified on a surface  of two complex
dimensions ${\cal S}$ resulting in a  four-dimensional field theory on $R^{7,1}\ra R^{3,1}
\times {\cal S}$,  with reduced global symmetry dictated by the decomposition
\[ SO(7,1)\times U(1)_R\ra SO(3,1)\times SO(4)\times U(1)_R \cdot \]
The 8-d spinor $\Psi_+$   decomposes as
\[\left(S_+,{\small\frac 12}\right)\ra \left((2,1),(2,1),\frac 12\right)\oplus \left((1,2),(1,2),-\frac 12\right)\]
with respect to $SO(4)\times U(1)_R\sim SU(2)\times SU(2)\times U(1)_R$, and similarly for $\Psi_-$.

The compact surface ${\cal S}$  is a nontrivial Riemannian four manifold and
spinors (needed to define the 4-d supercharges) are globally not  well defined. To preserve
${\cal N}=1$ SUSY we embed $U(1)_R$ into the $U(2)\in SO(4)$. (Note that the
K\"ahler structure of ${\cal S}$ is preserved only by one $U(2)$, so spinors can have
well defined properties only under the latter). Indeed, denoting  $J$  the generator of
$U(1)\in U(2)$  and $R$ that of $U(1)_R$, either of the combinations $J_{\pm}=J\pm 2 R$
preserves one supersymmetry. For the $\Psi_+$ and the $\epsilon_+$ generator choosing $J_+$ we have
\[\left(S_+,{\small\frac 12}\right)\ra \left\{(2,1)\otimes 2_1\right\}\oplus
\left\{(1,2)\otimes (1_2\oplus 1_0)\right\}\]
and analogously for $\Psi_-,\epsilon_-$.
The fields descending from $\Psi_+$ decomposition are denoted  as follows
\[\{(1,2)\otimes 1_0\}\ra\bar\eta^{\dot\alpha},\;\{(2,1)\otimes 2_1\}\ra\psi_{\bar m}^{\alpha},\;
\{(1,2)\otimes 1_2\}\ra\bar\chi_{\bar m\bar n}^{\dot\alpha},
\]
and constitute a zero, one and two form respectively,
\[\eta^{\dot\alpha},\;\psi^{\alpha}=\psi_{\bar m}^{\alpha}d\bar z^{\bar m},\;
\chi^{\dot\alpha}=\bar\chi_{\bar m\bar n}^{\dot\alpha}d\bar z^{\bar m}\wedge d\bar z^n\]
 and analogously for the conjugate $\Psi_-$.

Under $SO(3,1)\times U(2)\times U(1)_{J_+}$  the scalars $\varphi,\bar\varphi$
 transform as
\[\varphi =\{(1,1)\otimes 1_{-2}\},\;\bar\varphi =\{(1,1)_{+2}\}\]
while from the dimensional reduction of the 8-d vector $A$, we obtain  the 4-d
vector $A_{\mu}$ and the scalars $A_m(A_{\bar m})$  which transform as
\[A_{\bar m}=\{(1,1)\otimes 2_{+1}\} \cdot \]
The above fields pair up into one gauge multiplet and two ${\cal N}=1$ chiral multiplets as follows:
\begin{align}
(A_{\mu},\eta^{\alpha}),\;(A_{\bar m},\psi_{\bar m}^{\alpha}),\;(\varphi_{mn},\chi_{mn}^{\alpha})\cdot
\label{multi}
\end{align}

\section{KK-modes and the GUT scale}

 In F-theory, threshold corrections  associated to KK-massive modes
arise from gauge fields as well as from matter fields in the intersections.
As already  asserted in the introduction, KK-massive modes from
the chiral and the Higgs sectors add up to a common shift of the gauge coupling
constants at $M_{GUT}$. Indeed, we will prove that this happens  when the charges $q_i$ associated
to the matter curves $\Sigma_{q_i}$ are genuinely embedded into the function
${\cal T}(q_i)$ which defines the torsion.  Thus,  in this respect
the  F-theory case looks pretty much the same as in M-theory~\cite{Friedmann:2002ty}.
In the following, we will first give a brief account
 of the gauge thresholds computations adopting the techniques
of~\cite{Friedmann:2002ty} developed for $G_2$-manifolds, while
 we will follow~\cite{Donagi:2008kj} for the case of F-theory
we are interested in. Next, we will proceed with the computation of KK-thresholds from the
chiral matter and the Higgs curves.

\subsection{The gauge multiplet}

We write the decomposition of the $SU(5)$ gauge multiplet under the SM symmetry as
\[24\ra R_0+R_{-5/6}+R_{5/6}\]
with
\ba
R_0=(8,1)_0+(1,3)_0+(1,1)_0,\;R_{-5/6}=(3,2)_{-5/6},\;R_{5/6}=(\bar 3,2)_{5/6} \cdot
\label{24Ri}
\ea
 Massless fields  in the bulk are given by a topologically invariant quantity, the Euler characteristic $\cal X $, thus, in order to avoid the massless exotics  $R_{\pm 5/6}$ we impose the condition ${\cal X}(S,L^{5/6})=0$.
On the other hand, the massive modes in representations (\ref{24Ri})  induce threshold effects
to the running of the gauge couplings. At the one-loop level we write
\be
\label{gauge}
\frac{16\pi^2}{g^2_a(\mu)}=\frac{16\pi^2
k_a}{g^2_s}+b_a\log\frac{\Lambda^2}{\mu^2}+\mathcal{S}_a^{(g)},\quad
a=3,2,Y\,\cdot
\ee
Here  $\Lambda$ is the gauge theory cutoff scale, $k_a=(1,1,5/3)$ are the
normalization coefficients for the usual embedding of the Standard
Model into $SU(5)$,  $g_s$ is the value of the gauge coupling at the high  scale
and $\mathcal{S}_a^{(g)}$ stand for the gauge fields thresholds.
The one-loop $\beta$-function coefficients $b_a$ for the massless spectrum
(in the notation of~\cite{Friedmann:2002ty}) are
\be
\label{beta}
b_a=2\,\textrm{Str}_{M=0}Q^2_a\left(\frac{1}{12}-\chi^2\right)
\ee
where $\chi$ is the helicity operator  and $Q_a$ stands for the three
generators of the Standard Model gauge group $SU(3)\times SU(2)\times U(1)_Y$.
In computing the supertrace $\textrm{Str}$ we count bosonic contributions with weight $+1$
and  fermionic with $-1$. Similarly,
the one-loop threshold corrections from the KK-massive modes in $R_i$ are
\begin{align}
S_a^{(g)}&=2\,\sum_{i}\textrm{Tr}_{R_i}Q^2_a \,\,\,\textrm{Str}_{M\neq 0}\left(\frac{1}{12}-\chi^2\right)\log\frac{\Lambda^2}{M^2}\label{KKg} \cdot
\end{align}
The  squared masses of the KK modes in the threshold formula correspond to the massive spectrum
of the  Laplacian $\Delta_{k,R_i}$  acting on each $k$-form of the representation $R_i$.
In the previous section we saw that the spectrum (\ref{multi})  consists of zero, one and two form multiplets.
Each eigenvector of the zero-form Laplacian $\Delta_{0,R_i}$ contributes a vector multiplet with
helicities $1,-1,\frac 12,-\frac 12$, while the one-form Laplacian $\Delta_{1,R_i}$ gives a chiral multiplet with helicities $0,0,\frac 12,-\frac 12$. Finally,  $\Delta_{2,R_i}$ is associated to anti-chiral multiplets. The  sum of all the contributions  to the gauge fields thresholds  is
\be
\label{KK-contr3}
\mathcal{S}_a^{(g)}=2\sum_{i}{\rm Tr}_{R_i} (Q_a^2){\cal K}_i
\ee
with~\cite{Donagi:2008kj}
\be
\label{KK-contr4}
{\cal K}_i=\frac{3}{2}\log \det{'}\frac{\Delta_{0,R_i}}{\Lambda^2}-\frac 12\log \det{'} \frac{\Delta_{1,R_i}}{\Lambda^2}
-\frac 12\log \det{'} \frac{\Delta_{2,R_i}}{\Lambda^2}
\ee
where the prime  on det$'$ means that zero modes are omitted. Using the well known properties
 characterizing  the massive spectra of the Laplacians $\Delta_{k,R_i}$, it has
been shown~\cite{Donagi:2008kj} that  expression  (\ref{KK-contr4}) is the
 Ray-Singer analytic torsion ${\cal T}_i$~\cite{Ray:1973sb}; more precisely,
\be
\label{KK-contr4A}
2{\cal T}_{i}={\cal K}_i=2\log \det{'}\frac{\Delta_{0,R_i}}{\Lambda^2}-\log \det{'} \frac{\Delta_{1,R_i}}{\Lambda^2} \cdot
\ee
Note that for the trivial representation $R_0$ there exist zero-modes and
the torsion  differs from  ${\cal K}_{0}$  by a scaling
dependent part $\propto 2\log(V_S^{1/2}\Lambda^2)$ where $V_S$ is the volume of
the compact surface $S$. A detailed analysis on the scaling dependence can be found
in~\cite{Donagi:2008kj}. Returning to  (\ref{KKg}) we compute the traces  and use
the fact that ${\cal K}_{5/6}={\cal K}_{-5/6}$  to get
\begin{equation}
\label{S56}
\left({\cal S}_Y^{(g)},{\cal S}_2^{(g)},{\cal S}_3^{(g)}\right)=
\left(\frac{50}{3}{\cal K}_{5/6},\,6{\cal K}_{5/6}+4{\cal K}_0,\,4{\cal K}_{5/6}+6{\cal K}_0\right) \cdot
\end{equation}
Using the torsion ${\cal T}_i$ and the $\beta$-functions $b_a^{(g)}=(0,-6,-9)$, we can rewrite the above as
\begin{equation}
\label{S56_2}
{\cal S}^{(g)}_a=
\frac 43 b_a^{(g)} \left({\cal T}_{5/6}-{\cal T}_0\right)+20\,k_a{\cal T}_{5/6} \cdot
\end{equation}
Absorbing the term  proportional to $k_a$ into  a redefinition of  ${g}_s$
 we may now write the one loop equation (\ref{gauge}) for the running
 of the gauge couplings~\cite{Leontaris:2011pu} as
\begin{equation}
\label{gauge2}
\begin{split}
\frac{16\pi^2}{g^2_a(\mu)}
&=\left(\frac{16\pi^2 }{g^2_s}+20\,{\cal T}_{5/6}\right)k_a+
b_a^{(g)}\log\frac{\exp\left[4/3\left({\cal T}_{5/6}-{\cal T}_0 \right)\right]}{\mu^2 V_S^{1/2}}
\end{split} \cdot
\end{equation}
 The form (\ref{gauge2}) suggests that we  can define $M_{GUT}$ as~\cite{Leontaris:2011pu}
\begin{equation}
\label{MGUT}
M_{GUT}^2=\frac{\exp\left[4/3\left({\cal T}_{5/6}-{\cal T}_0 \right)\right]}{V_S^{1/2}}
\end{equation}
and a  gauge coupling $g_U$ at the GUT scale shifted by
\begin{equation}
\label{gredef}
\frac{16\pi^2 }{g^2_U}=\frac{16\pi^2 }{g^2_s}+20\,{\cal T}_{5/6} \cdot
\end{equation}
Furthermore, if we associate the world volume factor $V_S^{-1/4}$ with the characteristic
F-theory compactification scale $M_C$, we can write this equation as follows
\begin{equation}
\label{MGUT1}
M_{GUT}=e^{2/3\left({\cal T}_{5/6}-{\cal T}_0 \right)}\,M_C\,\cdot
\end{equation}
Thus, as far as  the gauge fields thresholds are concerned, we find that $M_{GUT}$
is given in terms of $M_C$ through an elegant relation. In the next section we will
present  the matter fields contributions and investigate the  conditions under which this
relation continues to hold true.

\subsection{The chiral matter}

Here, we will discuss  contributions arising from  chiral matter, the Higgs
fields and the possible exotic representations.  In F-theory constructions, these fields arise  in
the intersections of the GUT-brane with other 7-branes as well as from the decomposition
of the adjoint representation in the bulk. We have already imposed the conditions which avoid the exotic bulk zero modes $R_{-5/6}=(3,2)_{-5/6}$ and $R_{5/6}=(\bar 3,2)_{5/6}$, so we are only left with light matter fields at the intersections. In the  $SU(5)$ case, these correspond to the standard  $10,\overline{10}$ and $5,\bar 5$ non-trivial representations and contribute to the RG running   a term  of the form $b_a^x\log{{\Lambda'}^2}/{\mu^2}$ where $b_a^x$ are the $\beta$-function coefficients for the matter fields, and $\Lambda'$ a cutoff scale which may differ from the gauge cutoff $\Lambda$.

We should mention that the $U(1)_Y$-flux introduced in order to break   $SU(5)$ might eventually lead to incomplete $SU(5)$ representations, spoiling thus the gauge coupling unification.  However, it is still possible to work out realistic cases~\cite{Leontaris:2010zd,Dudas:2010zb,Leontaris:2011pu} where the matter fields add up to complete $SU(5)$ multiplets, so that the $b_a^x$-functions contribute in proportion to the coefficients $k_a$. Then, as in the case of the gauge contributions discussed earlier, we can absorb the logarithmic  ${\Lambda'}$-dependence into a redefinition of the gauge coupling. Nevertheless, the color triplet pair descending from the $5_H+\bar 5_H$ Higgs quintuplets must receive a mass at a  relatively high scale $M_X\le M_{GUT}$ so to avoid rapid proton decay. Taking all into account, we write (\ref{gauge2})
 in the form
\begin{equation}
\label{gauge3}
\begin{split}
\frac{16\pi^2}{g^2_a(\mu)}&=k_a\frac{16\pi^2}{g^2_{GUT}} +
(b_a^{(g)}+b_a)\log\frac{M_{GUT}^2}{\mu^2}+b^T_a\log\frac{M_{GUT}^2}{M_X^2}
\end{split}
\end{equation}
where we have split $b_a^x=b_a+b_a^T$ with $b_a$ denoting the MSSM $\beta$-functions  and $b_a^T$
 the color triplet part.

In the context of F-theory constructions, in addition to the light degrees of freedom on matter curves, one also has  to include contributions from Kaluza-Klein massive modes. As  already explained, this is in contrast to the case of $G_2$ manifolds, where no new contributions are introduced to the gauge coupling running apart from the massless states~\cite{Friedmann:2002ty}.
Threshold contributions  arise from the massive states along the $\Sigma_{\bar 5}$ and $\Sigma_{10}$ matter curves. To compute them we write down  the decompositions of the corresponding representations
\[
 10\ra (3,2)_{\frac 16}+(\bar 3,1)_{-\frac 23}+(1,1)_1,\quad
 {\bar 5}\ra (\bar 3,1)_{\frac 13}+(1,2)_{-\frac 12} \cdot
 \]
For each of the above matter curves we consider  the Laplacian acting on the  representations with eigenvalues corresponding to chiral and anti-chiral fields.
Thus, for the massive modes of $\Sigma_{10}$ we have
\[{\cal K}_{\Sigma_{10}}=-\frac 12\log{\rm det}'\frac{\Delta_{0,Y}}{{\Lambda'}^2}
-\frac 12\log{\rm det}'\frac{\Delta_{1,Y}}{{\Lambda'}^2}\]
and similarly for the $\Sigma_{\bar 5}$. Denoting by  $S_{a=3,2,Y}$ the thresholds to the
three gauge factors of the SM, for a representation $r$  we then have
 \[S_a^r=\sum_i2{\rm Tr}(Q_{a,r}^2){\cal K}_i \cdot \]
Computing the traces  we readily find  the KK-thresholds shown in  Table~\ref{Table}.
\begin{table}
\begin{center}
\begin{tabular}{lclc}
Thresholds&$SU(3)$& $SU(2)$& $U(1)$ \\
\hline

$S_a^{\bar 5}$ &  ${\cal K}_{1/3}$ & ${\cal K}_{-1/2}$&${\cal K}_{-1/2}+2/3\,{\cal K}_{1/3}$ \\
$S_a^{10}$ &  $2{\cal K}_{1/6}+{\cal K}_{-2/3}$ & $3{\cal K}_{1/6}$&$1/3{\cal K}_{1/6}+2{\cal K}_1+8/3{\cal K}_{-2/3}$ \\
\end{tabular}
\caption{Threshold corrections $S_a^{\bar 5}$ ,$S_a^{10}$ to the three gauge couplings  from Kaluza-Klein massive modes along the matter curves. }
\label{Table}
\end{center}
\end{table}

We will now further elaborate on the form of the corrections, and attempt
to recast them as a sum of two different pieces,  one being
proportional to $k_a$. The KK-thresholds induced by the $\bar 5$ can be written as follows;
\ba
S_a^{\bar 5}=-\frac{2}{3}\,\beta_a^{\bar 5}\,({\cal K}_{-1/2}-{\cal K}_{1/3})+k_a\cdot( {\cal K}_{-1/2})
\label{5cor}
\ea
where we have introduced the ``$\beta$''-coefficients
\[\beta_{3,2,1}^{\bar 5}=\{\frac 32,0,1\}\]
and, as usually, $k_a=(1,1,5/3)$.
For the $\Sigma_{10}$ we can write the thresholds related  to $U(1)_Y$ in the form
\ba
S_1^{10}&=&\frac 13\,{\cal K}_{1/6}+\frac 83{\cal K}_{-2/3}+2 {\cal K}_1\nn\\
&=&\frac 83\left({\cal K}_{-2/3}-{\cal K}_{1/6}\right)-2 \left({\cal K}_{1/6}-{\cal K}_1\right)
+\frac{15}3 {\cal K}_{1/6} \cdot
\ea
We observe that in the two parentheses  the $U(1)_Y$ charge differences obey the relation
 $q_i-q_j=-\frac 56$. This suggests that a non-trivial line bundle structure could be sought
 with the `periodicity'
 property ${\cal K}_{q_i}-{\cal K}_{q_j}=f(q_i-q_j)$ so that
\[ {\cal K}_{1/6}-{\cal K}_1={\cal K}_{-2/3}-{\cal K}_{1/6} \cdot \]
Adopting this assumption, we finally get
\ba
S_1^{10}&=&\frac{2}{3}\,({\cal K}_{-2/3}-{\cal K}_{1/6})+\frac 53\cdot(3\, {\cal K}_{1/6})\nn\\
S_2^{10}&=&0\,({\cal K}_{-2/3}-{\cal K}_{1/6})+1\cdot(3\, {\cal K}_{1/6})\nn\\
S_3^{10}&=&1\,(K_{-2/3}-K_{1/6})+1\cdot(3\, K_{1/6})\nn \cdot
\ea
These relations can be written in  compact form in straight analogy with (\ref{5cor}) as
\[S_a^{10}=\frac{2}{3}\beta_a^{10}({\cal K}_{-2/3}-{\cal K}_{1/6})+k_a\cdot(3\, {\cal K}_{1/6})\]
with $\beta_a^{10}=\beta_a^{\bar 5}$.

Recalling the Ray-Singer torsion ${\cal  T}_i$ we may write threshold terms for both matter
curves as follows
\ba
S_a^{\bar 5}&=&-\frac{4}{3}\,\beta_a^{\bar 5}\,({\cal T}_{-1/2}-{\cal T}_{1/3})+k_a\,( 2\cdot{\cal T}_{-1/2})
\label{TC5}
\\
S_a^{10}&=&+\frac{4}{3}\,\beta_a^{10}\,({\cal T}_{-2/3}-{\cal T}_{1/6})+k_a\,(6\cdot {\cal T}_{1/6})
\label{TC10}\cdot
\ea
 We now observe  that the hypercharge assignments in both $\Sigma_{10}$ and
 $\Sigma_{\bar 5}$   satisfy the same condition $q_i-q_j=-\frac 56$.
 Given this property and the fact that the torsion is a topologically invariant quantity,
 one could assume the existence of bundle structures for $\Sigma_{10}$ and $\Sigma_{\bar 5}$
matter curves characterized by the same topological properties so that
  we may envisage a specific embedding of the hypercharge generator
implying
\be
{\cal T}_{-1/2}-{\cal T}_{1/3} ={\cal T}_{-2/3}-{\cal T}_{1/6}=0\label{condition} \cdot
\ee
 In this limit, threshold contributions which are not proportional to $k_a$
cancel in both  $\Sigma_{10}$ and  $\Sigma_{\bar 5}$ curves.

In general, matter curves accommodating different representations of the
gauge group do not necessarily bear the same bundle structure. In particular,
in the case of $SU(5)$ it often happens that the  $\Sigma_{\bar 5}$  curve is of higher
genus than the $\Sigma_{10}$ for example. One of course could not
exclude the possibility that the
condition (\ref{condition}) can be separately  satisfied
for surfaces of different genera.  However, we mention
that in the recent literature one
 can find several examples where  $\Sigma_{10}$ and $\Sigma_{\bar 5}$
 curves are of the same genus and the required property holds true.
To give further support to our argument, we will briefly present
a model discussed in ref~\cite{Beasley:2008kw}.
Bearing in mind that in order to decouple gauge dynamics from gravity and allow
for the possibility $M_{GUT}\ll M_{Planck}$,  we choose
the  surface $S$ to be one of the del Pezzo type $dP_n$ with $n=1,2,\dots 8$.
We choose  $dP_8$  which is generated by the hyperplane divisor $H$ from
$\mathbb{P}^2$ and the exceptional divisors $E_{1,...,8}$ with intersection numbers
 \ba
 H\cdot H=1,\;H\cdot E_i=0,\;E_i\cdot E_j=-\delta_{ij}\cdot
 \ea
We also note that the canonical divisor  for $dP_8$ is
\ba
K_S&=&-c_1(dP_8)=-3H+\sum_{i=1}^8E_i \cdot
\ea
Then, denoting with $C$ and $g$ the class and the genus of a matter  curve respectively, we have
$C\cdot (C+K_S)=2 g-2$.
In the particular example of section 17 in ref~\cite{Beasley:2008kw} the $10_M$
chiral matter of the three generations resides on one $\Sigma_{10}$, with $C=2 H-E_1-E_5$
and the three $\bar 5_M$  on a single  $\Sigma_{5}^1$ curve with $C=H$. Higgs fields
$5_H$ and $\bar 5_{\bar H}$ are localized on different  $\Sigma_{5}^{2,3}$ matter curves
with classes $C= H-E_1-E_3$ and $ H-E_2-E_4$ respectively. Checking the
relevant  intersections,
one readily finds that all the above matter curves are of the same genus $g=0$ and
therefore the criterion is fulfilled.

Returning to the threshold contributions (\ref{TC5},\ref{TC10}), once the parts proportional
  to $\beta_a^{\bar 5},\beta_a^{10}$ cancel out we observe that the remaining contributions
  from KK thresholds are just those proportional to the coefficients $k_a$ and consequently,
  they only induce a shift of the gauge coupling value at $M_{GUT}$.
We finally get
\begin{equation}
\label{FinMG}
\begin{split}
\frac{16\pi^2}{g^2_a(\mu)}&=\left(\frac{16\pi^2 }{g^2_s}+20{\cal T}_{5/6}+6{\cal T}_{1/6}+2{\cal T}_{1/3}\right)k_a+
(b_a^{(g)}+b_a)\log\frac{M_{GUT}^2}{\mu^2 }
+b_a^T\log\frac{M_{GUT}^2}{M_X^2 }
\end{split} \cdot
\end{equation}
Thus, matter thresholds leave the GUT scale $M_{GUT}$ intact,  their only net
effect amounts to a further shift of the common gauge coupling. The value of the latter at the
GUT scale  is defined  by
\begin{equation}
\label{gGUT}
\frac{16\pi^2}{g^2_{GUT}}=\frac{16\pi^2 }{g^2_s}+20{\cal T}_{5/6}+6{\cal T}_{1/6}+2{\cal T}_{1/3} \cdot
\end{equation}
Note in passing that in the case where  KK-modes from the gauge multiplet are associated
to a  bundle with different properties, we denote ${\cal T}_{5/6}\ra {\cal  T}'_{5/6}$
 while the above analysis still holds.

We observe that (\ref{FinMG})  are just  the one-loop renormalization group equations for
the minimal $SU(5)$ GUT, with  extra color triplets  becoming massive at a scale $M_X\le M_{GUT}$.
We further note that in $F$-theory constructions, a $U(1)_Y$ flux mechanism is employed to
break the   $SU(5)$ symmetry,  inducing a splitting of the gauge couplings at the GUT
scale. Interestingly, this gauge coupling splitting is still consistent with a unification scale $M_{GUT}\sim 2\times 10^{16}$ GeV provided that the triplets receive a mass  at a scale determined by
 consistency conditions~\cite{Blumenhagen:2008aw,Leontaris:2009wi}.

\subsubsection{Example: The case of non-trivial line bundle}

In this section we will present an example of $\Sigma_{10},\Sigma_{\bar 5}$ matter curves
 with non-trivial  structure.  In particular, we will consider  the case of genus $g=1$
 Riemann surfaces  and use the torsion results of \cite{Ray:1973sb}
 to compute the KK-matter contributions.
We are interested in the masses of the KK  modes, that is the eigenvalues of the Laplacian
on a complex one dimensional Riemann surface.  Thresholds from these KK-massive modes are given as
functions of the torsion  which is expressed in terms  of the eigenvalues  through the zeta function associated to  the  Laplacian $\Delta_k$
\be
\label{Laplacian}
\Delta_{k,R(V)}=(\bar\partial+\bar\partial^{\dagger})^2=\bar\partial\bar\partial^{\dagger}+
\bar\partial^{\dagger}\bar\partial \, \cdot
\ee
If we  collectively denote  $\psi_k^{n}$ as the $k$-form eigenfunction, then
\be
\label{Eig}
\Delta_{k,R(V)}\psi_k^{n}=\lambda_n^k\psi_k^{n}
\ee
where $\lambda_n^k$ is the corresponding eigenvalue which in four dimensions
corresponds to a mass squared. The associated zeta-function  is given by
\ba
\zeta_{\Delta_k}(s)&=&\sum_n\frac{1}{\lambda_n^s}
          \;=\;\frac{1}{\Gamma(s)}\int_0^{\infty}t^{s-1}{\rm Tr}\left(e^{-\Delta_k\,t}\right)t
          \label{zetaD}
          \ea
 so that
\[\ln({\rm Det}\Delta_k)=-\left.\frac{d\zeta_{\Delta_k}(s)}{ds}\right|_{s=0} \cdot \]
 The torsion is written as
\ba
{\cal T}&=&\sum_k(-1)^{k+1}\,k\, \left.\frac{d\zeta_{\Delta_k}(s)}{ds}\right|_{s=0}\cdot
\label{zetp}
\ea
For our application,  we have already assumed a Riemann surface of genus $g= 1$ and a
character given by $\chi =\exp\{2\pi\,i(mu+nv)\}$ with the identification
$\chi \leftrightarrow u-\tau v$.
The eigenvalues are
\ba
\lambda_n&=&\frac{4\pi^2}{{\rm Im}{\tau}}\left|u+m-\tau (v+n)\right|^2 \cdot
\label{eig}
\ea
The eigenfunctions are
\[\psi_n=\exp\left\{\frac{2\pi i}{{\rm Im}\tau}{\rm Im}[z(u+m-\bar\tau(v+n))]\right\} \cdot \]

Given the eigenvalues (\ref{eig}), the torsion can be computed~\cite{Ray:1973sb}
using (\ref{zetp}) and (\ref{zetaD}). Because of its central role in this example, we present the basic
steps of its derivation, adapting the notation~\cite{Ray:1973sb} into our formalism.
Let us assume that $\tau=\tau_1+i\tau_2 $ and let us define $S_1={\rm Tr}\left(e^{-\Delta_k\,t}\right)$ which amounts to the calculation of the following double sum:
\begin{equation}
S_{1}=\sum\limits_{m,n=-\infty}^{\infty}\exp\left[  -\frac{4\pi^{2}t}{\tau
_{2}^{2}}\left(  \left(  u+m\right)  ^{2}+\tau^{2}\left(  v+n\right)
^{2}-2\tau_{1}\left(  u+m\right)  \left(  v+n\right)  \right)  \right] \cdot
\end{equation}
 Applying the Poisson summation formula we get
\begin{equation}
S_{1}=\frac{\tau_{2}}{4\pi t}\sum\limits_{m,n=-\infty}^{\infty}\exp\left[
-\frac{1}{4t}\left(  m^{2}\tau^{2}+n^{2}+2\tau_{1}mn\right)  +2\pi i\left(
mu+nv\right)  \right]  \cdot
\end{equation}
Putting  $a=\left(  m^{2}\tau^{2}+n^{2}+2\tau_{1}mn\right)$
and substituting into (\ref{zetaD}),  we get
\begin{equation}
\zeta\left(  s\right)  =\frac{\tau_{2}}{4\pi}\frac{1}{\Gamma\left(  s\right)
}\sum\limits_{m,n=-\infty}^{\infty}\int\limits_{0}^{\infty}dt~t^{s-2}%
e^{-\frac{a}{4t}}\exp\left[  2\pi i\left(  mu+nv\right)  \right]  \cdot
\end{equation}
For $s>1$ the integration gives
\begin{equation}
\zeta\left(  s\right)  =\frac{\tau_{2}}{4\pi}\frac{\Gamma\left(  1-s\right)
}{\Gamma\left(  s\right)  }\sum\limits_{m,n=-\infty}^{\infty}\left(  \frac
{4}{a}\right)^{1-s}\exp\left(  2\pi i\left( mu+nv\right)  \right)\cdot
\end{equation}
We readily now find that
\begin{equation}
\zeta'\left(  0\right)  =\frac{\tau_{2}}{\pi}\sum\limits_{m,n=-\infty
}^{\infty}\frac{\exp\left[  2\pi i\left(  mu+nv\right)  \right]  }{\left(
m^{2}\tau^{2}+n^{2}+2\tau_{1}mn\right)  }\ \cdot
\end{equation}
According to Kronecker's second limit theorem, the singular term $m=0$, $n=0$ has to be omitted~\cite{SingerSiegel}.
This way we get
\ba
\zeta'\left(  0\right)  =\frac{\tau_{2}}{\pi}\sum\limits_{n\neq0}\frac
{\exp\left[  2\pi inv\right]  }{n^{2}}+\frac{\tau_{2}}{\pi}\sum\limits_{m\neq
0}e^{2i\pi mu}{\displaystyle\sum\limits_{n=-\infty}^{\infty}}
\frac{e^{2i\pi nv}}{m^{2}\tau^{2}+n^{2}+2\tau_{1}mn}\ \cdot\label{zetapr}
\ea
The first sum is~\cite{Hansen 17.2.8}
\[
\sum\limits_{n\neq0}\frac{\exp\left[  2\pi inv\right]  }{n^{2}}=2\sum
\limits_{n=1}^{\infty}\frac{\cos2\pi vn}{n^{2}}=\frac{ 3\left(
2\pi v\right)  ^{2}-6\pi\left(  2\pi v\right)  +2\pi^{2}}{6}=2\pi
^{2}\left(  v^{2}-v+\frac{1}{6}\right)
\]
where $0<v<1$. The  $n$ sum in the second term of (\ref{zetapr})
 can be evaluated by means of the Poisson formula
 \begin{equation}
\sum\limits_{n=-\infty}^{\infty}f\left(  -n\right)  =\sum\limits_{n=-\infty
}^{\infty}\int\limits_{-\infty}^{\infty}e^{2\pi inx}f\left(  x\right)  dx
\end{equation}
where we take
\begin{equation}
 f(x)=\frac{e^{2i\pi vx}}{m^{2}\tau^{2}+x^{2}+2\tau_{1}mx} \cdot
\end{equation}

The denominator can be written as
\begin{equation}
m^{2}\tau^{2}+x^{2}+2\tau_{1}mx=\left(  m\tau_{1}+x\right)  ^{2}+m^{2}\tau
_{2}^{2}%
\end{equation}
so that
\ba
I& =&\int\limits_{-\infty}^{\infty}dx\frac
{e^{2i\pi\left(  n+v\right)  x}}{\left(  m\tau_{1}+x\right)  ^{2}+m^{2}%
\tau_{2}^{2}}=\int\limits_{-\infty}^{\infty}dx\frac{e^{-2i\pi\left(  n+v\right)  m\tau_{1}%
}e^{2i\pi\left(  n+v\right)  x}}{x^{2}+m^{2}\tau_{2}^{2}}\nn\\& =&\pi
\frac{e^{-2i\pi\left(  n+v\right)  m\tau_{1}}e^{-2\pi\left\vert v+n\right\vert
\left\vert m\tau_{2}\right\vert }}{\left\vert m\tau_{2}\right\vert } \cdot
\ea
Restricting to the upper plane so that $\tau_{2}=\operatorname{Im}\tau>0$, we finally get
\begin{gather*}
\zeta'\left(  0\right)  =2\pi\tau_{2}\left(  v^{2}-v+\frac{1}{6}\right)
+{\displaystyle\sum\limits_{n=-\infty}^{\infty}}
\sum\limits_{m\neq0}\frac{1}{\left\vert m\right\vert }\,e^{  -2\left\vert
m\right\vert \left\vert v+n\right\vert \pi\tau_{2}-2i\pi\left(  n+v\right)
m\tau_{1}\ +2i\pi mu } \cdot
\end{gather*}
The sum over $m$ gives
\begin{gather}
\sum\limits_{m\neq0}\frac{1}{\left\vert m\right\vert }\,e^{
-2a\pi\left\vert m\right\vert +2i\pi bm} =
-\ln\left(  1-e^{  -2\pi\left(  a+bi\right)  }  \right)
-\ln\left( 1-e^{ -2\pi\left(  a-bi\right)}  \right)
\end{gather}
or
\begin{gather*}
\zeta'\left(  0\right)  =2\pi\tau_{2}\left(  v^{2}-v+\frac{1}{6}\right)
-{\displaystyle\sum\limits_{n=-\infty}^{\infty}}
\ln\left\vert 1-e^{  -2\left\vert v+n\right\vert \pi\tau_{2}%
+2i\pi\left(  n+v\right)  \tau_{1}\ -2i\pi u}  \right\vert ^{2} \cdot
\end{gather*}
Consider now the exponent
\begin{equation}
2i\pi\left[  \left\vert v+n\right\vert i\tau_{2}+\left(  n+v\right)  \tau
_{1}-u\right]  \cdot
\end{equation}
For $n=0$ the terms inside the bracket become
$-\left(  u-\tau v\right)$ while
for $n>1$ we get
\begin{equation}
\left(  v+\left\vert n\right\vert \right)  i\tau_{2}+\left(  \left\vert
n\right\vert +v\right)  \tau_{1}-u=\left\vert n\right\vert \tau-\left(  u-\tau
v\right) \cdot
\end{equation}
For $n<-1$ we get
\ba
\left(  \left\vert n\right\vert -v\right)  i\tau_{2}+\left(  -\left\vert
n\right\vert +v\right)  \tau_{1}-u&=&-\left\vert n\right\vert \tau^{\ast}-\left(  u-v\tau^{\ast
}\right)=\left[  2i\pi\left(  \left\vert n\right\vert \tau+\left(  u-\tau v\right)
\right)  \right]  ^{\ast}\nn \cdot
\ea
All the above cases can be represented in a compact form as follows:
\[
\zeta'\left(  0\right)  =2\pi\tau_{2}\left(  v^{2}-v+\frac{1}{6}\right)  -%
{\displaystyle\sum\limits_{n=-\infty}^{\infty}}
\ln\left\vert 1-e^{ 2i\pi\left(  \left\vert n\right\vert
\tau-\varepsilon_{n}\left(  u-\tau v\right)  \right) } \right\vert
^{2}%
\]
where we have introduced the sign convention
\begin{equation}
\varepsilon_{n}=\mathrm{sign}\left(  n+\frac{1}{2}\right)  \cdot
\end{equation}

Now consider the function
\begin{equation}
g\left(  w,\tau\right)  =\prod\limits_{n=-\infty}^{\infty}\left(
1-\exp\left[  2i\pi\left(  \left\vert n\right\vert \tau-\varepsilon
_{n}w\right)  \right]  \right)  \cdot
\end{equation}
Separating out the zero mode we may write
\begin{equation}
g\left(  w,\tau\right)  =\left(  1-\exp\left[  -2i\pi w\right]  \right)
\prod\limits_{n=1}^{\infty}\left(  1-\exp\left[  2i\pi\left(  n\tau-w\right)
\right]  \right)  \prod\limits_{n=1}^{\infty}\left(  1-\exp\left[
2i\pi\left(  n\tau+w\right)  \right]  \right)  \cdot
\end{equation}
Using the nome $q=e^{i\pi\tau}$ we get
\begin{equation}
g\left(  w,\tau\right)  =2i\sin\pi w~e^{-i\pi w}\prod\limits_{n=1}^{\infty
}\left(  1-2q^{2n}\cos2\pi wu+q^{4n}\right)  \cdot
\end{equation}
The elliptic function $\vartheta_{1}$ is defined as%
\begin{equation}
\vartheta_{1}\left(  w,\tau\right)  =2q^{\frac{1}{4}}\sin\pi w\prod
\limits_{n=1}^{\infty}\left(  1-2q^{2n}\cos2\pi w+q^{4n}\right)  \left(
1-q^{2n}\right)  \cdot
\end{equation}
Using the Dedekind eta function
\begin{equation}
\eta\left(  \tau\right)  =q^{\frac{1}{12}}\prod\limits_{n=1}^{\infty}\left(
1-q^{2n}\right)
\end{equation}
we deduce that
\begin{equation}
\vartheta_{1}\left(  w,\tau\right)  =-iq^{\frac{1}{6}}e^{i\pi w}\eta\left(
\tau\right)  g\left(  w,\tau\right)
\end{equation}
or
\begin{equation}
\vartheta_{1}\left( w,\tau\right)  =-ie^{i\pi\left(  w+\frac{\tau}%
{6}\right)  }\eta\left(  \tau\right) g\left(  w,\tau\right) \cdot
\end{equation}

This way,
\ba
{\displaystyle\sum\limits_{n=-\infty}^{\infty}}
\ln\left\vert 1-e^{  2i\pi\left(  \left\vert n\right\vert
\tau-\varepsilon_{n}\left(  u-\tau v\right)  \right) }  \right\vert
^{2}&=&
\ln \left\vert \frac{\vartheta_{1}\left(  u-\tau v,\tau\right)  }%
{\eta\left(  \tau\right)  }\right\vert ^{2}\nn\\&& +\ln\left(  e^{-i\pi
\left(  u-\tau\left(  v-\frac{1}{6}\right)  \right)  }e^{i\pi\left(
u-\tau^{\ast}\left(  v-\frac{1}{6}\right)  \right)  }\right)\nn\\
&=&2\ln\left\vert \frac{\vartheta_{1}\left(  u-\tau v,\tau\right)  }{\eta\left(
\tau\right)  }\right\vert +\ln\left( e^{-2\pi\tau_{2}\left(  v-\frac{1}%
{6}\right)  }\right)\nn\\
&=&2\ln\left\vert \frac{\vartheta_{1}\left(  u-\tau v,\tau\right)  }{\eta\left(
\tau\right)  }\right\vert -2\pi\tau_{2}\left(  v-\frac{1}{6}\right)  \cdot
\ea
Finally, collecting all the terms we get
\ba
\zeta'\left(  0\right) &=&2\pi\tau_{2}\left(  v^{2}-v+\frac{1}{6}\right)
-2\ln\left\vert \frac{\vartheta_{1}\left(  u-\tau v,\tau\right)  }{\eta\left(
\tau\right)  }\right\vert +2\pi\tau_{2}\left(  v-\frac{1}{6}\right)\nn\\
&=&-2\ln\left\vert e^{i\pi\tau v^{2}}\frac{\vartheta
_{1}\left(  u-\tau v,\tau\right)  }{\eta\left(  \tau\right)  }\right\vert
\ea
Therefore, the analytic torsion is
\ba
{\cal T}_z=\ln\left|\frac{e^{\pi\,i\,v^2\tau}\vt_1(z,\tau)}{\eta(\tau)}\right|,\;z=u-\tau\,v \,\cdot
\ea

In order to use this result, we need to make a proper identification of  the hypercharge $q_i$. Let us first
recall the following identity for  $\vt_1(z,\tau)$:
\ba
\vt_1(z+\tau,\tau)&=&-e^{-\pi\,i\tau}e^{-2\pi\,i z}\vt_1(z,\tau) \,\cdot
\ea
For $z=u-\tau v$ this becomes
\ba
\vt_1(u-\tau v+\tau,\tau)&=&-e^{\pi\,i (2v-1)}e^{-2\pi\,i u}\vt_1(u-\tau v) \,\cdot
\label{ID}
\ea

In terms of the variables $u,v$, we observe that the transformation is essentially equivalent to the shift $v\ra v-1$,
i.e. the left part can be rewritten as $\vt_1(u-\tau (v-1),\tau)$. Consequently, for two different points $v,v-1$ the torsion reads
\ba
{\cal T}_{v}\equiv {\cal T}_{z=u-\tau v}&=&\ln\left|\frac{e^{\pi\,i\tau\,v^2}\vt_1(u-\tau v,\tau)}{\eta(\tau)}\right|
\label{TR1}\\
{\cal T}_{v-1}\equiv {\cal T}_{z=u-\tau (v-1)}&=&\ln\left|\frac{e^{\pi\,i\tau\,(v-1)^2}\vt_1(u-\tau( v-1),\tau)}{\eta(\tau)}\right| \cdot \label{TR2}
\ea
Using the identity (\ref{ID}) the numerator in the logarithmic quantity (\ref{TR2})
becomes
\ba
 e^{\pi\,i\tau\,(v-1)^2}\vt_1(u-\tau( v-1),\tau)&=&- e^{\pi\,i\tau\,(v-1)^2}e^{\pi\,i\tau (2v-1)}e^{-2\pi\,i u}\vt_1(u-\tau v)  \nn\\
&=&-e^{-2\pi\,i u}e^{\pi\,i\tau\,v^2}\vt_1(u-\tau v,\tau) \,\cdot
\ea
Now, substituting into the torsion formula and taking into account that $u$ is real, we obtain
\ba
{\cal T}_{z=u-\tau (v-1)}&=&\ln\left|-e^{-2\pi\,i u}e^{\pi\,i\tau\,v^2}\vt_1(u-\tau v,\tau)\right|
\nn
\\&=&\ln\left|e^{\pi\,i\tau\,v^2}\vt_1(u-\tau v,\tau)\right|
={\cal T}_{z=u-\tau v} \cdot
\ea
Considering now two successive hypercharge values $q_i,q_j$ such that $|q_i-q_j|=\frac 56$ and using the association
\ba
v_i&=&\frac{q_i}{|q_i-q_j|}
\label{Yv}
\ea
we get the identification
\[{\cal T}_{u-\tau v_i}\leftrightarrow{\cal T}_{q_i} \cdot \]

With this embedding we can easily see that the differences ${\cal T}_{-2/3}-{\cal T}_{1/6}$
and ${\cal T}_{-1/2}-{\cal T}_{1/3}$  vanish and the result (\ref{FinMG}) is readily obtained.

We stress that this example, although not fully realistic (since we have restricted our investigation to the flat torus) is sufficient to support  the aforementioned ideas.
In proposing the above identification we relied on the assumption that a $U(1)$ symmetry is naturally associated  with the one cycle of the torus,  while the hypercharge identification  seems to be in accordance with the notion of $U(1)$ fluxes piercing the matter curves. Indeed, we know that when the $U(1)$ fluxes are turned on they affect the multiplicity of the various massless representations along the matter curves. For example, assuming the $\Sigma_{\bar 5}$ matter curve, the number of   $5$'s and/or $\bar 5$'s is determined by the fluxes of $U(1)_i$'s  corresponding to some Cartan generators of the commutant gauge group inside $E_8$ (here being $SU(5)_{\perp}$). Furthermore, $U(1)_Y\in SU(5)_{GUT}$ determines in a similar manner the  splitting of the standard model representations obtained from the decomposition of $10$ and $\bar 5$'s. Indeed, in the presence of $U(1)_Y\in SU(5)_{GUT}$  flux, we can express  for example the splitting of the massless spectrum for $n$ units of hyperflux  for $5\ra (3,1)_{1/3}+(1,2)_{-1/2}$ as $\# (3,1)_{1/3} -\# (1,2)_{-1/2}=(v_d-v_l) n=n$.
 We notice that eq. (\ref{eig}) and  the hypercharge  association assumed in (\ref{Yv}) imply also  the same $v$-dependence of the corresponding massive modes.

\subsubsection{On matter curves with higher genera}

In the previous sections we have presented simple examples where
threshold corrections from KK states associated to genus one matter curves
do not alter the unification scale.
For $g=1$ the properties of the determinants are well understood and (at least
in the case of flat torus) we can corroborate our assumption for the $U(1)_Y$
embedding with an explicit  computation. However, in F-theory, we deal quite often  with
examples involving matter curves of higher genera  ($g\ge 2$).  In this case
a natural extension of the $\bar\partial$-torsion can be possibly related to the
Selberg's zeta function~\cite{Selberg}.
Then one has to deal with the rather non-trivial task of seeking specific
realistic cases where the required properties mentioned in the previous
sections are satisfied.  To convey an idea of the issues in this general case,
 we will give a brief account on the
possibility of implementing our analysis for $g>1$,  leaving a more detailed
consideration for future work.

To start with, we first note that the compact  Riemannian manifold
(for $g>1$) can be written as ${\cal H}/\Gamma$, that is,
 it can be identified as the quotient of the  upper half plane ${\cal H}$  by the group
of isometries $\Gamma$ of ${\cal H}$  with elements

\[\gamma =\left(\begin{array}{cc}a&b\\c&d\\ \end{array}\right)\in\Gamma:\;\ra \left(\begin{array}{cc}a&b\\c&d\\ \end{array}\right)z=
\frac{az+b}{cz+d}\]
with the condition $|a+d|>2$~\footnote{This is a space with hyperbolic geometry with metric $ds^2=y^{-2}(dx^2+dy^2)$ and constant negative curvature $R=-1$.}.
An element  $\gamma\in \Gamma$ is called primitive if it is not a power of
some other element in $\Gamma$.  An element $\gamma'$ is said to be
conjugate to another $\gamma$
 if there exists an element $\gamma_1$ in $\Gamma$ such that
\[\gamma'=\gamma_1\gamma\gamma_1^{-1}\]
We denote $\{\gamma\}$ the set of elements  which are conjugate
to $\gamma$.  This way, $\Gamma$ is the union of disjoint conjugacy classes.
If $\gamma_0$ is the primitive element of $\{\gamma\}$, then
any other element in the same class can be written as $\gamma =\gamma_0^n$
for some integer power $n$.  We mention that for a compact manifold
the element $\gamma\in \Gamma$ can also be written as
\[\gamma \in\Gamma:\;
\frac{z'-z_0}{z'-z_1}=e^{2\rho_{\gamma}}\,\frac{z-z_0}{z-z_1}\]
for two real fixed points $z_{0,1}$ and $\rho_{\gamma}>0$.
For given finite unitary representation $\chi(\gamma)$, the Selberg zeta-function
is defined~\cite{Ray:1973sb} as
\ba
Z(s,\chi)&=&\prod_{\{\gamma\}}\prod_{n=0}^{\infty}{\rm det}\left(1-\chi(\gamma) \, e^{-\rho_{\gamma}(s+n)}\right)
\label{SelbergZeta0}
\ea
with Re$(s)>1$. Hence, any required properties of the torsion could be  investigated
with respect to  its relation to the Selberg zeta function given by the  general formula
 (\ref{SelbergZeta0}).  For example, for two non-trivial unitary representations $\chi(\gamma)$ and $\chi'(\gamma')$  of $\Gamma$ and for a compact  Riemann surface of $g>1$, according to a theorem by Ray and Singer~\cite{Ray:1973sb} the difference $\ln({\cal T}_0(\chi))-\ln({\cal T}_0(\chi'))$  is proportional to $\ln(Z(\chi)-\ln(Z(\chi'))$. Several
 studies~\cite{Kean:72,D'Hoker:1986zt,Voros:1986vw,Steiner:1986qr,Elizalde:1997jv}
 have revealed interesting properties of Selberg's function. It is envisaged that one can
 find examples where the required  quantities exhibit periodicity properties and an appropriate hypercharge embedding could also be feasible. We plan to return to these issues in a future publication.

\section{Conclusions}

In unified theories emerging in the context of F-theory compactification, threshold
corrections from Kaluza-Klein massive modes  play a decisive role in gauge coupling
unification and the determination of the GUT scale.  In this work, we have revisited
this issue in the context of a specific minimal unification scenario, the F-theory
$SU(5)$ GUT.  Although the problem of KK thresholds is in general quite complicated, in
the model under consideration it gets remarkably simplified using the fact
that these thresholds  can be expressed in terms of a topologically invariant quantity, the
Ray-Singer  analytic torsion.  Previous considerations have shown that the KK-modes
from  the gauge multiplets can be absorbed into a redefinition of the effective GUT
mass scale and the string gauge coupling. However, the situation concerning
 KK-mode contributions emerging from the matter curves is less clear. Here,
 we have pursued  this issue one step further, and  analyzed  the conditions to be imposed on the
 matter spectrum and  the nature of bundle structure where matter resides, in
 order to ensure that the emerging  F-theory GUT  comply with low energy phenomenological
 expectations. We have given examples where matter resides on genus one matter
 curves with chiral matter forming complete $SU(5)$ multiplets, which
 are consistent with the minimal unification scenario. These models are also capable of reproducing the expected low energy values for the weak mixing angle and the strong gauge coupling.
 A short discussion is also devoted to the prospects of models possessing
  matter curves of higher genera.

\end{document}